\input harvmac
%%%%%%%%%%%%%%%%%%%%%%%
% preprint # in refs ?
\def\pre#1{ (preprint {\tt #1})}%use this to give preprint # in refs
%\def\pre#1{}%use this NOT to give preprint # in refs
%%%%%%%%%%%%%%%%%%%%%%%
\font\email=cmtt9
\def\d{{\rm d}}
\def\der{\partial}
\def\omslash{\raise.15ex\hbox{/}\kern-.57em\omega}
\def\aslash{\raise.15ex\hbox{/}\kern-.57em a}
\def\lamslash{\raise.15ex\hbox{/}\kern-.57em\lambda}
\def\Lslash{\,\raise.15ex\hbox{/}\mkern-13mu L}
\def\hslash{\raise.15ex\hbox{/}\kern-.57em h}
\def\Im{\mathop{\rm Im}\nolimits}
\def\Re{\mathop{\rm Re}\nolimits}
\def\tr{\mathop{\rm tr}\nolimits}
\Title{\vbox{\baselineskip12pt\hbox{RU-98-47}\hbox{math-ph/9810010}}}
{\vbox{\centerline{Adding and multiplying random matrices:}
\vskip2pt\centerline{a generalization of Voiculescu's formulae}}}
\centerline{P.~Zinn-Justin\footnote{$^\dagger$}{{\email
pzinn@physics.rutgers.edu}}}
\medskip\centerline{New High Energy Theory Center}
\medskip\centerline{126 Frelinghuysen Road,
Piscataway, NJ 08854-8019, USA}
\vskip .3in
\centerline{Abstract}
In this paper, we give an elementary proof of the additivity of the
functional inverses of the resolvents of large $N$ random matrices,
using recently developed matrix model techniques.
This proof also gives a very
natural generalization of these formulae to the case of measures
with an external field. A similar approach yields a
relation of the same type for multiplication of random matrices.
%\draft
\Date{10/98}
\newsec{Introduction}
In the theory of free random variables \ref\VOIC{D.V.~Voiculescu,
K.J.~Dykema and A.~Nica,
{\it Free random variables}, CRM monograph series (AMS, Providence RI, 1992).},
a remarkable additivity property of the functional inverses
of the spectral resolvents is found, allowing the addition of
random variables. There is also a
similar formula for multiplication of random variables. From now on,
we shall call them Voiculescu's formulae.
These mathematical results have some
interesting applications: indeed, it turns out that
large size (independent) random matrices with certain measures
are free variables. Therefore it becomes possible to compute the resolvent
of the sum of two random matrices from the knowledge of the
resolvent of the separate matrices, i.e. to add (and multipliy)
large random matrices.
This, in turn, applies to various physical situations:
``deterministic + random'' problem\nref\BHZ{E.~Br\'ezin, S.~Hikami
and A.~Zee, {\it Phys. Rev.} E51 (1995), 
5442\pre{hep-th/9412230}.}\nref\Zee{A.~Zee, {\it Nucl. Phys.} 
B474 (1996), 726\pre{cond-mat/9602146}.} [\xref\BHZ,\xref\Zee]
(noting that for gaussian randomness, the addition
formula essentially reduces to Pastur's equation
\ref\Pas{L.A.~Pastur, 
{\it Theor. Math. Phys. (USSR)} 10 (1972), 102.}), 
random matrix methods applied to QCD \ref\NPZ{M.A.~Nowak, 
G.~Papp and I.~Zahed,
{\it Phys. Lett.} B389 (1996), 137\pre{hep-ph/9603348}\semi
R.A.~Janik, M.A.~Nowak, G.~Papp, J.~Wambach and I.~Zahed,
{\it Phys. Rev.} E55 (1997), 4100\pre{hep-ph/9609491}\semi
R.A.~Janik, M.A.~Nowak, G.~Papp, J.~Wambach and I.~Zahed,
{\it Acta Phys. Polon.} B28 (1997), 2949\pre{hep-th/9710103}.},
non-hermitean random matrices \ref\FZee{J.~Feinberg and A.~Zee, {\it
Nucl. Phys.} B501 (1997), 643\pre{cond-mat/9704191}.}, the Anderson
model \ref\NS{P.~Neu and R.~Speicher, {\it J. Stat. Phys.} 
80 (1995), 1279\pre{cond-mat/9410064}.}.
Since in the ``planar'' large $N$ limit ($N$ size of the matrices) that we
consider here one cannot compute $n$-point connected correlations of the
eigenvalues or other $N\to\infty$ subdominant corrections,
alternative methods (such as the supersymmetric
method, see review \ref\GMGW{T.~Guhr, A.~Mueller-Groeling
and H.A.~Weidenmueller,
{\it Phys. Rept.} 299 (1998), 189\pre{cond-mat/9707301}.}
and references therein)
may be required for a more detailed analysis;
but for many problems, it is still very important to
be able to compute the density of eigenvalues ($1$-point function),
which Voiculescu's formulae provide.

It is therefore of great interest to find an elementary proof of these
formulae. We shall mention one such proof
by Zee \Zee\ of the addition formula, which is based on a perturbative
approach: the measures of two random hermitean matrices $M_1$ and
$M_2$ are assumed to be derived
from an action of the form $\tr V(M)$ where $V$ is a polynomial,
and the perturbative expansion is represented diagrammatically,
leading to a diagrammatic proof of Voiculescu's formula.

However, this proof has limitations. First it assumes
$U(N)$-invariance of the actions. Of course one might object
that if we assume both measures to be non $U(N)$-invariant,
then Voiculescu's formula is not true any more (there a
is an obvious counter-example, which is the case of two fixed
matrices). And if one measure is $U(N)$-invariant and the
other is not, one can freely replace the non-invariant measure
$\d\mu(M)$ with an invariant one by averaging on the unitary
group:
$$\d\mu_{\rm eff}(M)=\int_{\Omega\in U(N)}
\!\!\!\!
\d\Omega\,\, \d\mu(\Omega M\Omega^\dagger)$$
This replacement will not affect the resolvent of the sum of $M$
and of another random matrix with $U(N)$-invariant measure;
however, it is not completely innocent since even if the original
measure $\d\mu(M)$ was derived from a simple polynomial action,
there is no reason for $\d\mu_{\rm eff}(M)$ to possess the
same property.

We see that the problem is that this proof does not allow for
general enough
measures; in particular, a very interesting physical application
is the case of a fixed matrix (for the
deterministic + random problem),
where the corresponding measure
is highly singular ($\delta$ function) and does not fit
in this perturbative framework.

We propose in this paper a new proof of both
addition and multiplication formulae, which makes very few
assumptions on the measures; it is based on recently developed
matrix model
techniques \ref\PZJ{P.~Zinn-Justin, {\it Commun. Math. Phys.} 194
(1998), 631\pre{cond-mat/9705044}.}
which have been successfully applied to physical models
\ref\PZJK{V.A.~Kazakov and P.~Zinn-Justin, preprint {\tt
hep-th/9808043}.}. In section 2, we shall show how to add matrices
by introducing an external field in the measure (as in \PZJ); and in section
3, we shall multiply matrices by adding this time a character
in the measure (as in \PZJK). The proof
has the obvious advantage that it generalizes the
usual addition formula to the case of a measure with an
external field (and similarly, the multiplication formula
to the case of a measure with a character insertion).
Section 3 is devoted to a summary of the results and conclusions.

\newsec{Adding random matrices}
Before adressing the problem
of the addition of several matrices, we shall explain our approach
by considering a single $N\times N$ hermitean 
matrix $M$ with a $U(N)$-invariant measure $\d\mu(M)$.
The only assumption we
make about this measure is that the diagonalization of $M$ leads to
a saddle point for the eigenvalues of $M$; that is, 
after integrating out the angular degrees of freedom,
the dominant large $N$ contribution is obtained by simply considering
the eigenvalues to be fixed (up to a permutation of the
eigenvalues). This a reasonable assumption, since as $N\to\infty$,
there are only $N$ eigenvalues, as opposed to the $N^2$ degrees
of freedom of the full matrix. For example,
a typical measure that is encountered in physical problems is:
\eqn\defmeas{\d\mu(M)= \prod_i \d M_{ii}\, 
\prod_{i<j} \d\Re M_{ij}\, \d\Im M_{ij}\, \exp(-S(M))}
where $S(M)$ is the action, which is invariant --
$S(M)=S(\Omega M \Omega^\dagger)$ for all $\Omega\in U(N)$ --
and scales likes $N^2$ as $N\to\infty$, which ensures a saddle point
for the eigenvalues\foot{For example,
$S(M)$ can be of the form: $S(M)=N \tr V(M)$, 
where $V$ is a given polynomial; but more general
actions with products of traces are possible.}. However,
the action $\d\mu(M)$ does not have to be of the form \defmeas,
and in particular can be more singular\foot{For example,
for a
fixed matrix, after averaging over the unitary group $U(N)$, the
measure is a $\delta$ function for the eigenvalues.}.

We now introduce the partition function with an additional external
field $A$ (see 
\nref\EXT{E.~Br\'ezin and D.~Gross,
{\it Phys. Lett.} B97 (1980),120\semi
D.J.~Gross and M.J.~Newman, {\it Phys. Lett.} B266 (1991), 291.}
\nref\POT{V.A.~Kazakov, {\it Nucl. Phys.} B (Proc. Suppl.) 4 (1988), 93\semi
J.M.~Daul, preprint {\tt hep-th/9502014}.}
[\xref\EXT,\xref\POT,\xref\PZJ] for the appearance
of such an external field in physical models):
\eqn\defA{Z(A)=\int \d\mu(M) \exp(N\tr MA)}
where $A$ is a fixed hermitean matrix. When $N\to\infty$,
one must consider
a sequence of $N\times N$ matrices $A$ such that their spectral density
tends to a continuous density $\rho_A(a)$ on the real axis.
Since the measure is $U(N)$-invariant $Z(A)$ depends only on the
eigenvalues of $A$, and for definiteness, we shall choose $A$
to be diagonal, with eigenvalues $a_j$, $j=1\ldots N$. 

We can go over to the eigenvalues $\lambda_i$ of $M$
by using the Itzykson--Zuber--Harish
Chandra formula \ref\IZ{Harish~Chandra, {\it Amer. J. Math.} 79
(1957), 87\semi
C.~Itzykson and J.B.~Zuber, {\it J. Math. Phys.} 21 (1980), 411.}:
\eqn\Zeig{Z[a_j]=\int\d\mu[\lambda_i]
{\det[\exp(N\lambda_i a_j)]
\over\Delta[\lambda_i]\Delta[a_j]}}
where $\Delta[\cdot]$ is the Van der Monde determinant, 
and $\d\mu[\lambda_i]$ is the resulting measure
on the eigenvalues; for example, with a measure of the type \defmeas,
we have:
\eqn\Zeigb{Z[a_j]=\int\prod_i\d\lambda_i \exp(-S[\lambda_i])
\Delta[\lambda_i] {\det[\exp(N\lambda_i a_j)]
\over\Delta[a_j]}}
where we have used the fact that the action $S$ only depends on the 
eigenvalues $\lambda_i$ of $M$.

Finally we introduce the logarithmic
derivative of $Z$ with respect to the eigenvalues $a_j$. According
to \defA, it is simply expressed as an average
$${1\over N}{\der\over\der a_j} \log Z[a_j]=
\left< M_{jj} \right>_A$$
where the subscript $A$ indicates that the average
is made in the presence of
the external field, i.e. with the measure $\d\mu(M) \exp(N\tr MA)$.
A more useful expression for this logarithmic derivative is found
by applying \Zeig:
\eqn\logder{
{1\over N}{\der\over\der a_j} \log Z[a_j]=
{1\over N}\left< {\der\over\der a_j}
\log{\det[\exp(N\lambda_i a_j)]
\over\Delta[a_j]}\right>_A
}
The kind of derivative that appears
in \logder\ has been studied in \PZJ; we shall briefly
review the results we need, and refer the reader
to the appendix 1 of \PZJ\ for the technical details.
In the large $N$ limit, the spectral density of $A$ tends
by definition to the continuous density $\rho_A$, and similarly,
since there is a saddle point for the eigenvalues of $M$,
we assume that the spectral density $\rho_M$ of $M$ becomes also
continuous.
Then, the derivative with respect to $a_j$ \logder\ becomes an analytic
function $f(a_j)$ of its argument $a_j$, of the form:
\eqn\deff{f(a)=\lambda(a)-\omega_A(a)}
Let us define the two functions in \deff: $\omega_A(a)$ is
the resolvent of $A$:
$$\omega_A(a)={1\over N}\tr{1\over a-A}=\int{\d a'\rho_A(a')\over a-a'}$$
It is an analytic function of $a$ except 
for a cut on the support of $A$ (which is contained
in the real axis).
If we introduce the notation $\omslash_A(a)={1\over2} (\omega_A(a+i0)
+\omega_A(a-i0))$ for $a$ real, so that
$\omega_A(a\pm i0)=\omslash_A(a)\mp i\pi \rho_A(a)$,
then
\eqn\omsl{\omslash_A(a_j)
={1\over N}{\der\over\der a_j} \log\Delta[a_j]}

Similarly, $\lambda(a)$ is defined by the following requirements:
it has the same cut as $\omega_A(a)$ on the support of $\rho_A$, and it
satisfies:
\eqn\lsl{\lamslash(a_j)={1\over N}\left<{\der\over\der a_j} 
\log\det[\exp(N\lambda_i a_j)]\right>_A}
Of course, $\lambda(a)$ may have more cuts than $\omega_A(a)$, whose
positions are left undefined; so one should really think of
$\lambda(a)$ as a multi-valued function, living on a branched covering
of the complex plane.

Note that combining \omsl\ and \lsl\ and using the fact that
$\omega_A(a)$ and $\lambda(a)$
have the same cut, one finds the expression \deff\ for
the logarithmic derivative \logder.

It is now possible to connect the function $\lambda(a)$ with the
resolvent $\omega_M(\lambda)$ of $M$:
$$\omega_M(\lambda)=\left< {1\over N} \tr {1\over\lambda-M}\right>_A
=\int{\d\lambda' \rho_M(\lambda')\over\lambda-\lambda'}$$

Indeed, it was shown in \PZJ\ (see also the earlier work 
\ref\MAT{A.~Matytsin, {\it Nucl. Phys.} B411 (1994), 805\pre{hep-th/9306077}.})
that if one introduces
in a symmetric way the function $a(\lambda)$ with
the same cut as $\omega_M(\lambda)$ and such that
$$\aslash(\lambda_i)={1\over N}\left<{\der\over\der \lambda_i} \log\det[\exp(
N\lambda_i a_j)]\right>_A$$
then $a(\lambda)$ and $\lambda(a)$ are functional inverses of each
other as multi-valued analytic functions.

Let us now take the limit $A\to 0$\foot{Note that
if one directly takes $A=0$, expressions such as \omsl\ and \lsl\
become meaningless; so one must consider a limit
where the support of $\rho_A$ has a finite size but
becomes smaller and smaller.}, that is $\rho_A(a)\to \delta(a)$ or
still $\omega_A(a)\to 1/a$. In this limit, from the Itzykson--Zuber--Harish
Chandra formula, one infers that 
$\det[\exp(N\lambda_i a_j)]/\Delta[a_j]\sim \Delta[\lambda_i]$, 
so that
$a(\lambda)$ tends to the resolvent $\omega_M(\lambda)$,
Therefore, for $A=0$, $\lambda(a)$ is precisely the functional
inverse of the resolvent we were looking for.

It is now clear that the obvious factorization property
$$\exp(N\tr(M_1+M_2)A)
=\exp(N\tr M_1A)\exp(N\tr M_2A)$$
implies the additivity of the average of its logarithmic
derivative:
$$\left<(M_1+M_2)_{jj}\right>_A=
\left<M_{1;jj}\right>_A + \left<M_{2;jj}\right>_A$$
On condition that the two matrices $M_1$ and $M_2$ are independent,
this can be rewritten as the additivity of the function
$$\lambda(a)-\omega_A(a)$$
or for the particular case $A=0$:
$$\lambda(a)-{1\over a}$$
This is the essence of Voiculescu's formula for adding random matrices.

Let us see how this works more explicitly, by considering
two independent random matrices $M_1$ and $M_2$
with measures $\d\mu_1(M_1)$ and $\d\mu_2(M_2)$.
We shall assume both measures to be $U(N)$-invariant, even though,
as explained in the introduction, it is not more difficult to prove
the formula with only one $U(N)$-invariant measure and a non-invariant
one. Both measures are
such that there exists a saddle point for the eigenvalues of $M_1$ and
$M_2$.

Then one introduces the partition function with an external field:
\eqn\Ztwo{Z(A)=\int\!\!\int \d\mu_1(M_1) \d\mu_2(M_2) \exp(N\tr(M_1+M_2)A)}
Again, due to $U(N)$-invariance of both measures,
$Z(A)$ depends only on the eigenvalues $a_j$ of $A$. Therefore
we can write that
\eqn\aver{
Z[a_j]=\int_{\Omega\in U(N)}\!\!\!\!\d\Omega\, Z(\Omega A\Omega^\dagger)
}
where we use the normalized Haar measure on $U(N)$.
By performing explicitly the integration over $\Omega$ (once more, the
Itzykson--Zuber--Harish Chandra integral), we immediately obtain that
$$Z[a_j]=\int\!\!\int\d\mu_1(M_1)\d\mu_2(M_2) {\det[\exp(N\lambda_i a_j)]\over\Delta[\lambda_i]
\Delta[a_j]}$$
where the $\lambda_i$ are the eigenvalues of $M_1+M_2$. We can now introduce
the usual logarithmic derivative with respect to $a_j$, which is of the form
\eqn\Ztwob{{1\over N}{\der\over\der a_j} \log Z[a_j] 
={1\over N}\left< {\der\over\der a_j}
\log{\det[\exp(N\lambda_i a_j)]
\over\Delta[a_j]}\right>_A
=\lambda(a_j)-\omega_A(a_j)}
where $\lambda(a)$ is connected with the matrix $M_1+M_2$; in particular,
for $A=0$, it is the functional inverse of the resolvent $\omega_{M_1+M_2}
(\lambda)$.

On the other hand, one can diagonalize separately $M_1$ and $M_2$,
since the partition function completely factorizes as $Z(A)=Z_1(A) Z_2(A)$,
with obvious notations. One finds
$$
Z[a_j]=\int\!\!\int \d\mu_1(M_1)\d\mu_2(M_2)
{\det[\exp(N\lambda_{1;i}a_j)]\over\Delta[\lambda_{1;i}]\Delta[a_j]}
{\det[\exp(N\lambda_{2;i}a_j)]\over\Delta[\lambda_{2;i}]\Delta[a_j]}
$$
Therefore
\eqn\Ztwoc{{1\over N}{\der\over\der a_j} \log Z[a_j] 
= (\lambda_1(a_j)-\omega_A(a_j))
+(\lambda_2(a_j)-\omega_A(a_j))}

Combining \Ztwob\ and \Ztwoc\ we find that the relation
\eqn\voic{\lambda_1(a)+\lambda_2(a)=\lambda(a)+\omega_A(a)}
is valid on the support of the density of $\rho_A(a)$, and
by analytic continuation is therefore valid on the whole complex plane.

For $A=0$, the functions $\lambda(a)$, $\lambda_1(a)$,
$\lambda_2(a)$ are functional inverses of the corresponding
resolvents, and $\omega_A(a)=1/a$, so that \voic\ reduces
to Voiculescu's formula for adding free variables.
However, the relation \voic\ still holds for arbitrary $A$,
thus generalizing Voiculescu's formula in a highly
non-trivial way.

\noindent{\it Remarks:}

1) if we assume that only one
measure (e.g. $\d\mu_1(M_1)$) is $U(N)$-invariant,
then $Z(A)$ no longer depends only on the eigenvalues
of $A$; but we can take \aver\ as a definition
of $Z[a_j]$, and then the rest of the proof works identically
(except that instead of diagonalizing $M_2$, one integrates
over $\Omega$).

2) In the $A=0$ case,
the connection to the usual diagrammatic interpretation
is the following. One can show that for $A=0$, our definition
of $f(a)=\lambda(a)-1/a$ is equivalent to:
$f(a)\equiv{1\over N} {\d\over\d a}
\log\left<\exp(Na M_{11})\right>$, where $M_{11}$ is an arbitrarily
chosen diagonal element.
$\left<\exp(NaM_{11})\right>$ being a generating function of the
moments of $M_{11}$, its $\log$ generates the connected moments:
$\log\left<\exp(Na M_{11})\right>=\sum_{n=0}^\infty {N^n\over n!}
a^n\left<M_{11}{}^n\right>_c$.
Furthermore,
using $U(N)$-invariance of the measure and both {\it planarity}
and {\it connectedness} of the diagrams that appear
in the perturbative expansion, one has the following
large $N$ equality:
$$\left<M_{11}{}^n\right>_c\buildrel N\to\infty\over\sim
{(n-1)!\over N^n} \left<\tr M^n\right>_c$$
Therefore one finds the usual expansion
$f(a)=\sum_{n=0}^\infty
a^n{1\over N}\left<\tr M^{n+1}\right>_c$.

\newsec{Multiplying random matrices}
The same type of argument applies to the multiplication of random
matrices. Let us start again with a single hermitean random matrix
with a measure $\d\mu(M)$ which leads to a saddle point on the
eigenvalues. We define the partition function {\it with a
character}:
\eqn\defZh{Z[H]={1\over\dim H}\int \d\mu(M) \chi_H(M)}
Here $H$ is a (holomorphic) irreducible representation of $GL(N)$; it can be
parametrized in the following way: $H=\{h_j;\, j=1\ldots N\}$,
where the $h_j$, $j=1\ldots N$, which
form a decreasing sequence of integers,
are the {\it shifted highest weights} of $H$ (the shifted
highest weights $h_j$ are connected with the usual highest
weights $m_j$ by the formula: $h_j=N-j+m_j$). $\chi_H(M)$
is the character of $H$ taken at $M$.
Using Weyl's formula for the character $\chi_H(M)$ and
the fact that $\dim H={\rm cst}\, \Delta[h_j]$, we can rewrite
$Z[H]$ in terms of the $h_j$:
\eqn\defZhb{Z[h_j]=\int\d\mu(M)
{\det[\lambda_i^{h_j}]\over\Delta[\lambda_i]\Delta[h_j]}}
This expression is very similar to Eq. \Zeig\ obtained after
use of the Itzykson--Zuber--Harish Chandra formula. It is now
clear that the same formalism will apply (see appendix 2 of \PZJ\ and
\PZJK\ for more details).

In the large $N$ limit, we assume that the $h_j/N$ (note the important
rescaling of a factor of $N$) tend to a continuous density
$\rho_H(h)$. We can then consider the $h_j/N$ as continuous real
variables,
and introduce the logarithmic derivatives
\eqn\logderh{{\der\over\der h_j} \log Z[h_j]=\left<{\der\over\der h_j}
\log{\det[\lambda_i^{h_j}]\over\Delta[h_j]}\right>_H}
(${\der\over\der h_j}$ stands for ${1\over N}{\der\over\der (h_j/N)}$)

We are now led to the introduction of two functions: the resolvent
$\omega_H(h)$
$$\omega_H(h)=\int {\d h'\rho_H(h')\over h-h'}$$
and the function $L(h)$ which has the same cut as
$\omega_H(h)$ and whose mean value on it is
$$\Lslash(h_j/N)=\left<{\der\over\der h_j}
\log\det[\lambda_i^{h_j}]\right>_H$$
We finally define $\lambda(h)=\exp L(h)$.

The eigenvalues also have a saddle point density $\rho_M(\lambda)$,
with its associated resolvent $\omega_M(\lambda)$, and there
is a function $h(\lambda)$ which satisfies
$$\hslash(\lambda_i)={1\over N}\left<\lambda_i{\der\over\der\lambda_i}
\log\det[\lambda_i^{h_j}]
\right>_H
$$
and which has the same cut as $\lambda\,\omega_M(\lambda)$.
$h(\lambda)$ and $\lambda(h)$ are of course functional inverses
of each other. Note that we were forced to introduce an extra
factor of $\lambda$ in the definition of $h(\lambda)$, which is
the crucial difference with the previous section. Indeed, let
us now choose $H$ to be the trivial representation, so that
$h_i=N-i$, that is
$$\omega_H(h)=\log {h\over h-1}$$
Then $\det[\lambda_i^{h_j}]=\Delta[\lambda_i]$ and therefore
$h(\lambda)=\lambda\,\omega_M(\lambda)$: $\lambda(h)$ is now
the functional inverse of $\lambda$ times the resolvent, and
not of the resolvent itself, which is something completely
different.

Let us now write down a formula for multiplying two matrices
$M_1$ and $M_2$, with associated measures $\d\mu_1(M_1)$ and
$\d\mu_2(M_2)$. As before, at least one of the two measures
must be $U(N)$-invariant. We introduce the partition function
with a character:
\eqn\Ztwoh{Z(H)=
\int\!\!\int \d\mu_1(M_1) \d\mu_2(M_2) {\chi_H(M_1M_2)\over\dim H}}
Direct application of the previous formalism to
the product $M_1 M_2$ leads to
\eqn\Ztwohb{{\der\over\der h_j} \log Z[h_j]= \log \lambda(h_j/N)
- \omega_H(h_j/N)}
where $\lambda(h)$ is associated to the product $M_1M_2$.

On the other hand, since one of the two measures is $U(N)$-invariant,
we can write that
$$Z[h_j]={1\over\dim H}
\int_{\Omega\in U(N)}\d\Omega
\int\!\!\int \d\mu_1(M_1) \d\mu_2(M_2) \chi_H(\Omega M_1\Omega^\dagger
M_2)$$
Using orthogonality relations for matrix elements of irreducible
representations, we can integrate over $\Omega$:
$$Z[h_j]=
\int\!\!\int \d\mu_1(M_1) \d\mu_2(M_2) {\chi_H(M_1)\over\dim H}
{\chi_H(M_2)\over\dim H}$$
The logarithmic derivative can now be written as:
\eqn\Ztwohc{{\der\over\der h_j} \log Z[h_j]= (\log \lambda_1(h_j/N)
- \omega_H(h_j/N))+(\log \lambda_2(h_j/N)-\omega_H(h_j/N))}
where $\lambda_1(h)$ and $\lambda_2(h)$ are the functions
associated in the usual way to the matrices $M_1$ and
$M_2$.

Comparing \Ztwohb\ and \Ztwohc\ and exponentiating the resulting
formula, as is more appropriate for multiplying matrices, we find:
\eqn\voich{\lambda_1(h)\lambda_2(h)=\lambda(h)\exp(\omega_H(h))}
that is the multiplicativity of the function $\lambda(h) \exp(-\omega_H(h))$.

If we now restrict ourselves to the case of the trivial
representation, $\lambda(h)$, $\lambda_1(h)$, $\lambda_2(h)$
are functional inverses of $\lambda$ times the corresponding
resolvents, and $\omega_H(h)=\log(h/(h-1))$ so that
\eqn\voichb{\lambda_1(h)\lambda_2(h)=\lambda(h){h\over h-1}}

Note once more that the functions $\lambda(a)$ in Eq. \voic\ and
the functions $\lambda(h)$ in Eq. \voich\ are not directly related
to each other since they are expressed in terms of different
variables.

\newsec{Conclusion}
We have proven two main formulae: Eq. \voic\ for the addition
of random matrices, and Eq. \voich\ for their multiplication.
As far as the author knows, the second formula,
even in its usual form (Eq. \voichb), does not
have a simple diagrammatic proof.

The proofs used above have the advantage that they clearly
highlight the key hypothesis needed for the results to hold:
i) $U(N)$-invariance of (at least one of) the two measures,
and ii) an analyticity property of the resolvents. Let us discuss
these two points.

\nref\VOICb{D.~Voiculescu,
{\it Invent. Math.} 104 (1991), 201.}\nref\Spe{R.~Speicher, {\it RIMS}
29 (1993), 731.}
The $U(N)$-invariance of the measure is an essential ingredient of
the proof: without it one cannot integrate over the unitary
group to use the Itzykson--Zuber--Harish Chandra formula
or the orthonality formula for characters. 
This is completely consistent with the assertion found
in the mathematical literature [\xref\VOICb,\xref\Spe]
that the two matrices
should be independently $U(N)$-rotated with respect to each other
in order to ensure freeness.
As has already been
mentioned, this hypothesis is obviously necessary (case of two fixed
matrices); but let us also note that when one keeps the
external field $A$ non-zero (or the representation $H$ non-trivial),
then one obtains addition/multiplication formulae which are
{\it different} from Voiculescu's formulae (and, generically,
incompatible with them); so that for these measures
(which of course also break
$U(N)$-invariance), the random matrices are no more free variables,
but still satisfy addition/multiplication formulae.

The analyticity property of the resolvents stems from the fact
that we have assumed the matrices to be hermitean, which
prevents the eigenvalues from moving freely in the complex plane,
and creating dense regions where the resolvent is no more analytic.
However, the proof does not really make use of the hermiticity of
the matrices, and the generalization to non-hermitean matrices
might provide some useful insight on these more complicated
matrix models.

\bigskip\centerline{\bf Acknowledgement}
This work was supported in part by the DOE grant DE-FG02-96ER40559.

%\vfill\eject
%\appendix{A}{Functional inversion relation}
%blahblah similar to appendix 1 of \PZJ.
%blahblah restricted version had already appeared in the literature
%(see in particular \ref\MAT{A.~Matytsin, {\it Nucl. Phys.} B411 (1994), 805.})
%blahblah\foot{This hypothesis may be violated -- this is the case
%when two cuts of $\lambda(a)$ overlap. Then the definition given in section
%1 does not work, and Eq. () should be
%taken as the correct definition of $\lambda(a)$.
%
\listrefs

\bye